\documentclass[conference]{IEEEtran}
\IEEEoverridecommandlockouts
\pdfoutput=1
\usepackage{multirow}
\usepackage{array}
\usepackage{units}
\usepackage[euler]{textgreek}
\usepackage{graphicx}
\usepackage{setspace}
\usepackage[font=normalsize]{subfig}
\usepackage{balance}

\begin{document}
	
	\title{
		Body Dust: Ultra-Low Power OOK     Modulation Circuit for Wireless Data Transmission 
		in Drinkable sub-100\textmu m-sized Biochips        
	}
	
	\author{
		\IEEEauthorblockN{Gian Luca Barbruni\IEEEauthorrefmark{1}\IEEEauthorrefmark{2}, Paolo Motto Ros\IEEEauthorrefmark{3}, Simone Aiassa\IEEEauthorrefmark{1}\IEEEauthorrefmark{2}, Danilo Demarchi\IEEEauthorrefmark{1}\IEEEauthorrefmark{2}, Sandro Carrara\IEEEauthorrefmark{2}}
		\IEEEauthorblockA{\IEEEauthorrefmark{1}Department of Electronics and Telecommunications, Politecnico di Torino, Turin, Italy}
		\IEEEauthorblockA{\IEEEauthorrefmark{2}Integrated Circuits Laboratory, \'{E}cole Polytechnique F\'{e}d\'{e}rale de Lausanne, Neuch\^atel, Switzerland}
		\IEEEauthorblockA{\IEEEauthorrefmark{3}Electronic Design Laboratory, Istituto Italiano di Tecnologia, Genova, Italy}
		\IEEEauthorblockA{Corresponding author: simone.aiassa@polito.it}
	}
	
	\maketitle

	\begin{abstract}
		
		In this paper, we discuss the feasibility of creating an UltraSound (US) communication circuit to wireless transmit outside the body diagnostic information from multiplexed biosensors chip built on the top layer of a drinkable CMOS Body Dust cube. The system requires to be small enough (lateral size of less than \unit[100]{\textmu m}) to mimic the typical size of a larger blood cell (diameter around \unit[30]{\textmu m} for white cells) and so support free circulation of the cube in human tissues. The second constraint came from the low-power consumption requirement, with the energy provided by an external US base station. The results of the feasibility study demonstrate the possibility to design an architecture for a data transmission transponder by using a \unit[0.18]{\textmu m} CMOS process, with sub-\unit[10]{\textmu W} of power consumption and a total chip area of \unit[43x44]{\textmu m\textsuperscript{2}}. The paper also discusses the limits of the designed system and the need for further improvements toward real applications in Body Dust diagnostics.
		
	\end{abstract}
	
	\section{Introduction}
	
	The Body Dust proposed approach~\cite{bodydust} has the target to develop highly innovative solutions for precision medicine applications, to obtain a more efficient and preventive diagnosis, for example in oncology. The system to be implemented is an ingestible micro-active sensing-chip with a wireless connection to an external base station, namely, through an UltraSound (US) communication system. The in-body biochip is conceived to answer by reflecting back a portion of the interrogating US wave by backscattering modulation. A similar approach has been followed by~\cite{lsk_implant} in which a US LSK modulator for deep implanted medical devices has been realized. In our work, the transducer model comes from~\cite{lsk_implant, pressure_touch}, because of its equal f0 operating frequency and its scalability in dimensions, required in the presented application. In the Body Dust system, the battery-less transponder retrieves a DC power supply by rectifying the incoming US signal. The literature suggests that through this method the power transfer achieves only a few \textmu W for the operation of the entire chip~\cite{lowpowerclock}, being power consumption a crucial point to determine tag performance. Considering the target application, the Body Dust chip must present a lateral size of less than \unit[100]{\textmu m}. Recent results in literature indicate that is possible to realize CMOS micro-systems at these sizes. In particular, a complete CMOS IC for multi-purpose sensing was realized in 2014~\cite{sara2014} with a total size of \unit[0.36]{mm\textsuperscript{2}}, while a \unit[2.5]{\textmu m} radius working electrode was realized for biosensing~\cite{microelectrode}, and a micro-coils with \unit[50]{\textmu m} of later size was published in 2016~\cite{microcoil}. This work presents a CMOS architecture to transmit the maximum number of information outside the body through On-Off Keying (OOK) modulation for backscattering. Similar transmissions have been proposed in 2013 for Neural Dust sensing approach~\cite{neural_dust}, where the system included three nodes, an external transceiver, a sub-dural one, and the proper Neural Dust device. Meanwhile, in our design, only two components are expected, which are the body dust tag and the external base station which works both as power transmitter and as data receiver. Moreover, \cite{neural_dust} is lacking a communication circuit and the impedance modulation is achieved through switching MOSFET only. Although in this work, we include a communication protocol to transmit outside the information. We base our design on an end-to-end, from the sensor read-out front-end to the transmitter back-end, event-driven digital architecture. The event-based approach aims in at pursuing an improved quality-energy trade-off~\cite{energy_trade_off} and has been already showed promising results in this specific application~\cite{low_power_architecture}. 
	
	\begin{figure}[!b]
		\centering
		\includegraphics[width=0.98\linewidth]{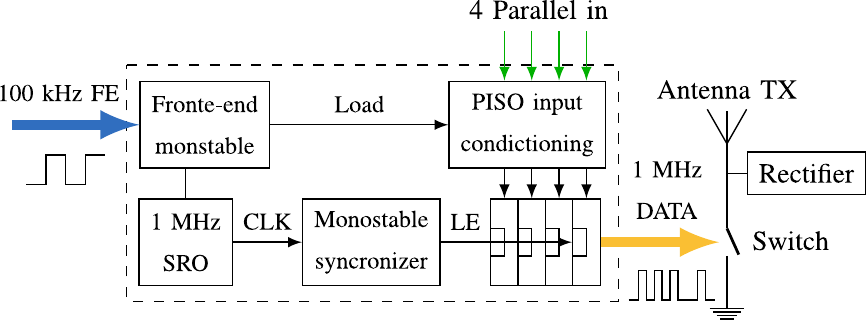}
		\vspace{.2cm}
		\caption{System block diagram.}
		\label{fig:fig1}
	\end{figure}
	
	\section{Proposed Architecture}
	The presented work here focuses on the design of the CMOS circuit for communication, highlighting both the resulting size and power consumption. The proposed system contains a \unit[1]{MHz} Starved Ring Oscillator (SRO), a front-end monostable, a monostable synchronizer, a 4-in Parallel-Input Serial-Output (PISO) register, and a modulator switch (Fig.~1). 
	The biosensor front-end is described in~\cite{bodydust}. A potentiostat on the top metal layer followed by an I-to-f (Current to frequency) converter, translate the biochemical information in a quasi-digital format. The quasi-digital signal is a digital signal carrying an analog information, already implemented for low power biosensing~\cite{quasi_digital_aiassa}, and it could directly trigger an event-driven wireless transmitter (e.g. an impulse radio~\cite{impluseradio}). This signal is considered here as the input of the proposed circuit. Since here we have to distinguish among different sensor front-end types, each event is labelled with a unique source identifier. This binary address is provided from the multiplexed layer of the chip that is designed to allow for at least 5 different quasi-digital signals (3 bits for the identification) to be transmitted. In~\cite{bodydust}, the I-to-F is designed with a sensitivity of \unit[7]{kHz/nA}, the sensitivity of the electrochemical sensor mounted on the top metal layer almost \unit[180]{nA/(mmol$\cdot$mm)}, and the estimated current range at the Counter Electrode is between \unit[1]{nA} and \unit[5]{nA}. 
	From this, the maximum frequency of the quasi-digital signal coming from the potentiostat is considered to be \unit[100]{kHz}.
	
	\begin{figure}[!b]	
		\centering
		\includegraphics[width=0.98\linewidth]{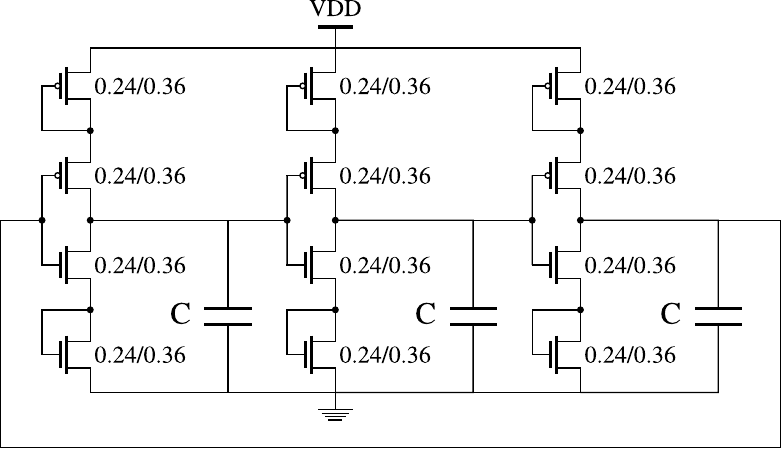}
		\caption{Three stages starved Ring Oscillator (RO).}
		\label{fig:fig2}
	\end{figure}
	  
	\subsection{The Starved Ring Oscillator (SRO)}
	
	The SRO generates the clock to synchronize the system and to modulate the serial output at a frequency of \unit[1]{MHz}. The oscillation frequency of the SRO is approximated by~(1)
	\begin{equation}
	f = \frac{I}{N \cdot C_{tot} \cdot V_{dd}}
	\end{equation}
	where N is the number of stages of the ring, C\textsubscript{tot} is the total capacitance, V\textsubscript{dd} the power supply (1.8 V) and I the current. The here implemented topology is a three-stages-starved-ring configuration with three equal capacitors, as shown in Fig.~\ref{fig:fig2}.
	
	The values of C and MOSFETs are optimized to guarantee a \unit[1]{MHz} oscillation, minimizing the capacitance to reduce the final size of the chip. The selected value for C is \unit[200]{fF}. By increasing the length of the transistors is possible to reduce the capacitance accordingly, but as a matter of fact, the total area improvement is less than \unit[10]{\%}. Further, this would negatively affect the SRO performance reliability by making the oscillation frequency highly dependent on the parasitic capacitance. After simulating different trade-off in this regard, we deemed the proposed design a safe and good solution. The SRO is tested through a post-layout simulation, sweeping the temperature from \unit[27]{\textsuperscript{o}C} to \unit[47]{\textsuperscript{o}C}, which is the typical range of biomedical application. Considering this range of temperature, the simulation results prove that the maximum shifting in frequency is \unit[10]{\%}. The performance of the designed oscillator are compared with similar works~\cite{adjustable, process_voltage} in Table~\ref{tab1}.

	\begin{table}[!t]
		\caption{SRO performance comparison.}
		\label{tab1}
		\centering
		\renewcommand{\arraystretch}{1.3}
		\begin{tabular}{lccc}
			\hline
			& \cite{adjustable} & \cite{process_voltage} & This work \\
			\hline
			CMOS technology &    \unit[130]{nm} & \unit[130]{nm} &    \unit[180]{nm} \\
			Voltage supply (V) &    3.3 &    3.3 &    1.8 \\
			External component &    Yes &    No &    No \\
			Frequency (MHz)    & 22.5-360 &    1 &    1 \\
			Power (\textmu W) &    1515\textsuperscript{*} &    87 &    9.5 \\
			Area (mm\textsuperscript{2}) &    0.05 &    0.054 &    0.00116 \\
			\hline
			\multicolumn{4}{l}{\textsuperscript{*}At 200 MHz}
		\end{tabular}
	\end{table}
	
	\begin{figure}[!b]
		\centering
		\includegraphics[width=0.9\linewidth]{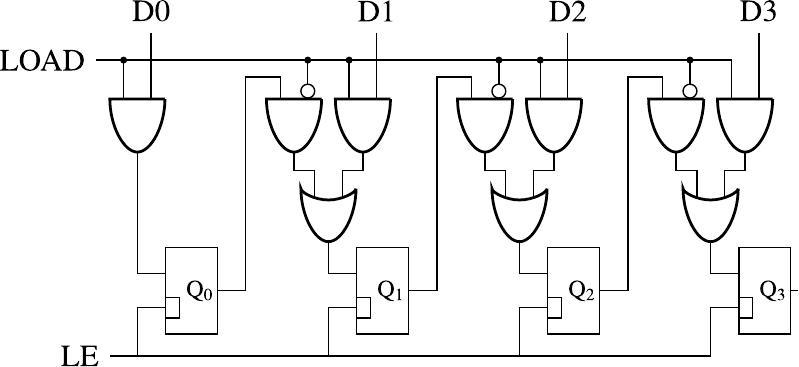}
				\vspace{.2cm}
		\caption{Four inputs PISO register.}
		\label{fig:fig3}
	\end{figure}

	\subsection{The Parallel-Input Serial-Output (PISO) register}
	
\begin{figure}[!t]
		\centering
		\includegraphics[width=0.7\linewidth]{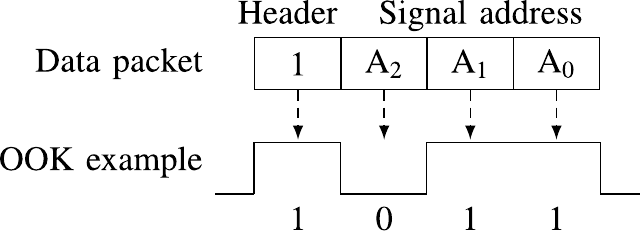}
		\vspace{.2cm}
		\caption{Packet template and example of OOK modulation. }
		\label{fig:fig4}
	\end{figure}

	The here implemented 4-input PISO register is composed of four D-latches, triggered by a Latch Enable (LE) that is generated by the monostable synchronizer, as represented in Fig.~\ref{fig:fig3}. Latches are preferred than Positive Edge Triggered (PET)-D Flip Flop in order to both improve the synchronism and reduce the size. 
	The inputs of the latches (parallel word) came from a simple combinational circuit in which the LOAD signal is triggered by the front-end monostable to control writing and shifting time. In particular, when LOAD is at the logic state high the word is written to each latch, when it is at the logic state low the PISO acts like a shift register. After four clock cycles, the serial data is ready to modulate the final MOSFET. Here the size of the transistors is the minimum provided from the selected Process Design Kit (PDK). By reducing the technology node it is possible to scale all the circuit minimizing the area. The transmitted data packet is encapsulated similarly to the work done for a wireless multi-channel event-driven tactile sensing glove~\cite{glove}. As shown in Fig.~\ref{fig:fig4}, the packet is composed by a header (bit 1 always) to declare the starting of the transmission and three bits for the signal address (A\textsubscript{2}, A\textsubscript{1}, A\textsubscript{0}). The minimum to distinguish the 5 signals treated by the sensor front end multiplexer.
	
	\subsection{The Front-end Monostable}
	
	The front-end monostable receives the Front-End (FE) signal and it is composed by a D-latch as delay block. Whenever a rising edge of the incoming signal is detected, a pulse is generated as the LOAD signal that triggers the PISO Input Conditioning in the writing or shifting operations. The LOAD signal retains its logic state high enough for almost \unit[1]{\textmu s} to detect the possible variations of the parallel word coming from the multiplexed layer and to ensure that at least one edge of the clock is asserted. 
	
	\subsection{The Monostable Synchronizer}
	
\begin{figure}[!t]
		\centering
		\includegraphics[width=0.98\linewidth]{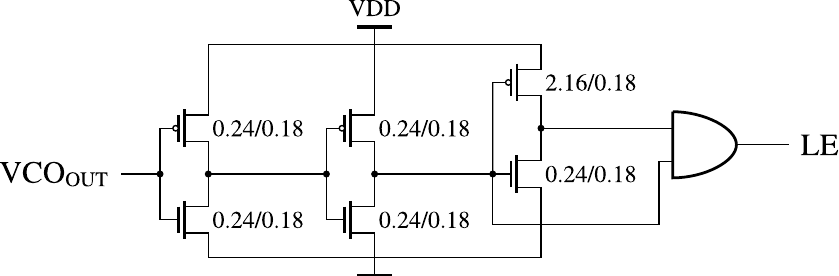}
		\vspace{.2cm}
		\caption{Implemented monostable synchronizer.}
		\label{fig:fig5}
	\end{figure}

	The monostable synchronizer generates the LE signal. It is a delay block made by a chain of inverters as presented in Fig.~\ref{fig:fig5}. The monostable synchronizer is the most critical part of the circuit since it has to be sized accordingly to the number of components receiving the trigger, namely, four latches in the PISO register and one latch in the front-end monostable. The output LE signal is a \unit[1]{ns} pulse over the activation threshold. To improve the synchronization, the monostable synchronizer is centred with respect to all the latch in the layout~\cite{rabaey}.
	
	\section{On-Off Keying (OOK) Modulator}
	
	\begin{figure}[!b]
		\centering
		\includegraphics[width=0.98\linewidth]{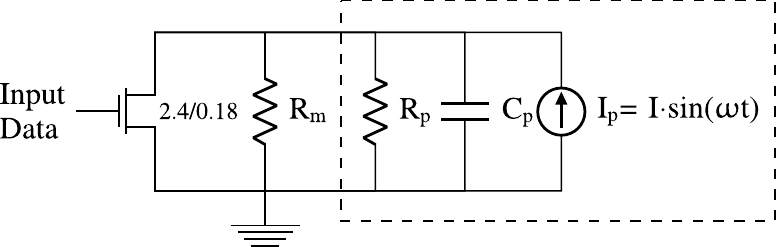}
		\vspace{.2cm}
		\caption{Electrical model of the piezo and incident wave (red box), R\textsubscript{M} modulator and NMOS switch.}		
		\label{fig:fig6}
	\end{figure}

	The final On-Off Keying (OOK) modulator is implemented by a MOSFET (NMOS in particular) acting as a switch. The drain driving current is high enough to modulate the impedance of the piezoelectric transmitting antenna. The size of the transistors is chosen accordingly to the transducer electrical equivalent impedance. In the designed circuit modulator resistance (R\textsubscript{m}) is set exactly equal in value to the resistance of the electrical model of the piezoelectric antenna~\cite{lsk_implant}: when the switch is open, R\textsubscript{m} is equal R\textsubscript{p}, and the full matched condition is achieved. On the contrary, when the switch is closed, R\textsubscript{m} is equal to R\textsubscript{on} and the mismatched condition is achieved. The circuit is shown in Fig.~6, R\textsubscript{m} and R\textsubscript{p} are \unit[3552]{$\Omega$}, C\textsubscript{p} is \unit[62.75]{pF} and the current I is \unit[10]{\textmu A}.

	\section{Simulation Results}
	
	\begin{figure}[!b]
		\centering
		\includegraphics[width=0.9\linewidth]{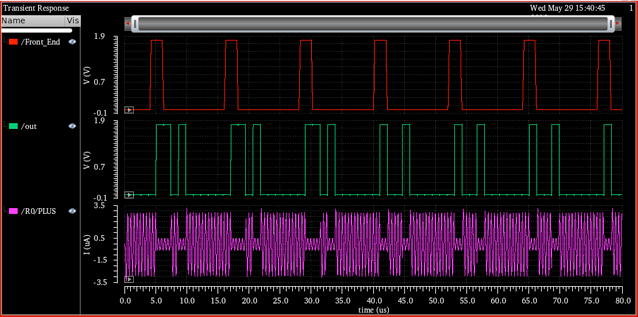}
		\caption{Full-circuit transient analysis with two different packet templates (1101 and 1001).}
		\label{fig:fig7}
	\end{figure}
	\begin{figure}[!b]
		\centering
		\includegraphics[width=0.9\linewidth]{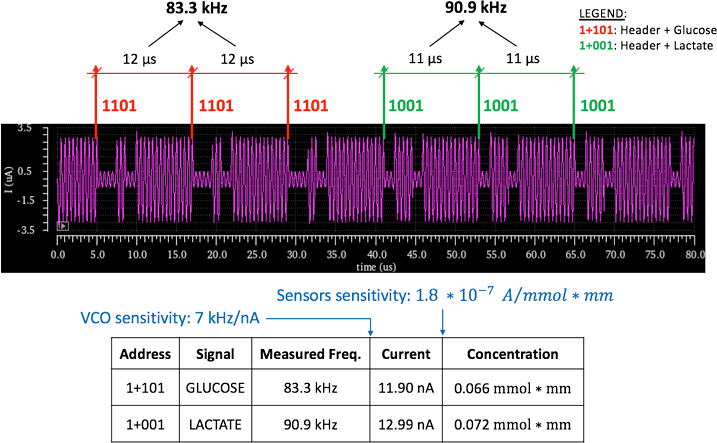}
		\caption{Different frequency for two different signals extracted and modulated.}
		\label{fig:fig8}
	\end{figure}
	
	To show the versatility of the proposed architecture, Fig.~7 presents the simulation of our circuit into a more complex system including the management of multiple sensors. Using two different templates (1101 and 1001) is possible to emulate the detection of two different molecules concentrations (for example glucose and lactate). 
	To simulate the activation of two biosensors on the analog front-end and the treatment of the signals provided by the multiplexing layer, the communication circuit continuously receives the quasi-digital signal from the same FE input without any information about the type of signal. The discriminant is the signal address (A\textsubscript{2}, A\textsubscript{1}, A\textsubscript{0}), provided through a counter by the multiplexer. In particular, the pink wave is received at the base station (assuming the absence of the tissues as losing medium) and the information is reconstructed as highlighted in Fig.~8. This proves the functioning of the circuit and, moreover, its scalability compared to more complex multi-sensor systems.
	In Fig.~9 the final layout of the proposed architecture is depicted: switch and modulator are not present since they require to be tuned accordingly to other transducer models. 
	The area breakdown of the designed architecture is shown in Fig.~10. Most of the area is occupied by the ring oscillator and the same is valid considering the power consumption. More than \unit[95]{\%} of the averaged power is consumed by the SRO. Nevertheless, post-layout simulation of the circuit results in the total power consumption of \unit[9.7]{\textmu W}. The instantaneous current required for the oscillation results to be \unit[350]{\textmu A/pulse} (\unit[630]{\textmu W/pulse}), while \unit[450]{\textmu A/pulse} (\unit[810]{\textmu W/pulse}) are spent for the total transmission. Table~II presents a comparison with~\cite{lsk_implant}, which demonstrates that the here proposed OOK modulation based implementation shows a \unit[80]{\%} reduction in the average power consumption, and a \unit[75]{\%} lower instant power consumption.

	\begin{figure}[!b]
		\centering
		\includegraphics[width=0.9\linewidth]{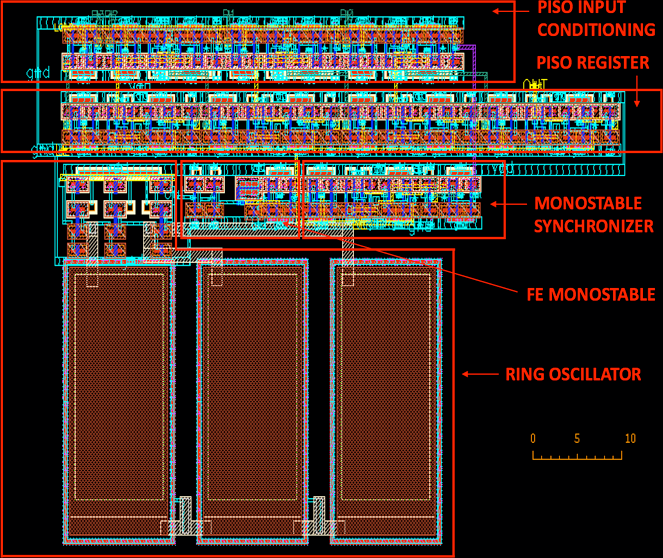}
		\caption{Implemented circuit layout with partition of the area for each block.}
		\label{fig:fig9}
	\end{figure}
	
	\begin{figure}[!b]
		\centering
		\includegraphics[width=0.9\linewidth]{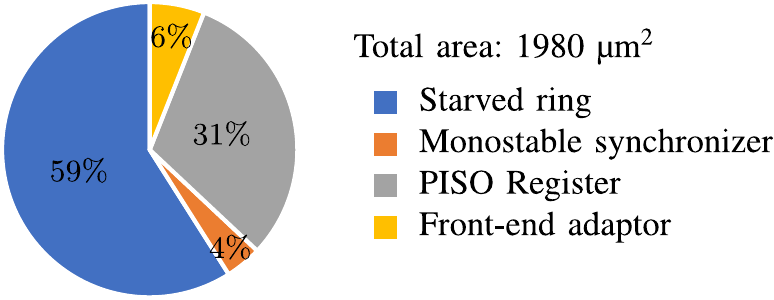}  
				\vspace{.2cm}
		\caption{Total area per block in the implemented circuit.}
		\label{fig:fig10}
	\end{figure}
	
	\section{Conclusion}
	
	In this paper, a novel architecture for data transmission in Body Dust diagnostics has been presented. The results of such feasibility study demonstrate the possibility to design an architecture for a US data transponder by using a \unit[0.18]{\textmu m} CMOS process, with an average power consumption of less than \unit[10]{\textmu W} and a total chip area of
	\unit[43x44]{\textmu m\textsuperscript{2}}. Even if we have implemented the design of a starved architecture intending to reduce both area and power consumption, the study shows that the source of major area consumption is still the ring-oscillator. Although, this is the first work to prove the feasibility of creating an architecture to wireless transmit data outside from the Body Dust cube. Additional work is still needed to further reduce the chip area, e.g. by considering smaller CMOS nodes. Even if our simulations already show promising results w.r.t ultrasound integrated communication devices, a smaller CMOS node together with subthreshold design approaches would help to further reduce the power consumption.
	
	\section*{Acknowledgment}
	
	The authors would like to thank Catherine Dehollain from EPFL, Lausanne, for useful discussions about power transmission with UltraSound.
	
	\begin{table}[!t]
		\caption{Modulation comparison.}
		\label{tab2}
		\centering
		\renewcommand{\arraystretch}{1.3}
		\begin{tabular}{lcc}
			\hline
			& \cite{lsk_implant} & This work \\
			\hline
			Modulaton &    FSK & ASK \\
			Average Power (\textmu W) & 49 & 9.7 \\
			Instant Power (mW/pulse)& 3.2 & 0.81\\
			\hline
		\end{tabular}
	\end{table}

	\balance

\end{document}